\documentclass{ws-ijmpd}
\UseRawInputEncoding
\usepackage[latin1]{inputenc}
\usepackage[super,compress]{cite}
\usepackage[unicode,pdftex]{hyperref}
\begin{document}

\markboth{Alexander F. Zakharov}
{Shadows near supermassive black holes: from a theoretical concept to GR test}

%
\catchline{}{}{}{}{}
%

\title{SHADOWS NEAR SUPERMASSIVE BLACK HOLES: FROM A THEORETICAL CONCEPT TO GR TEST
 }

\author{ALEXANDER F. ZAKHAROV
}


\address{
Bogoliubov Laboratory for Theoretical Physics, JINR,\\
141980 Dubna, Russia}


%

\maketitle

\begin{history}
\received{30 December 2022}
\revised{30 June 2023}
\end{history}

\begin{abstract}
General relativity (GR) passed many astronomical tests but in majority of them GR
predictions have been tested in a weak gravitational field approximation.
 Around 50 years ago a shadow has been introduced by J. Bardeen as a purely theoretical concept
but due to an enormous progress in observational and computational facilities this theoretical prediction
has been confirmed and the most solid argument for an existence of supermassive black holes in Sgr A* and M87* has been obtained.

\end{abstract}


\ccode{PACS numbers:04.20.-q, 04.30.Tv, 04.70.-s, 04.70.Bw, 04.80.Cc, 98.35.Jk}


\section{Introduction}

In 1915 A. Einstein created the general relativity (GR)
and the history of this remarkable achievement started since a great success because this theory helped to
explain old standing problem and new observational predictions have been done, really
 Einstein explained the anomaly of Mercury's motion, i.e. to solve the problem posed by Le Verrier in 1859, which
worried astronomers and celestial mechanics experts for more than half a century. In addition Einstein showed that
a deflection of light near point like source in two times larger than the Newtonian deflection of light.
In 1919 during a solar eclipse, it was found
that the deviations of the star positions near the solar disk  correspond to the predictions of the GR, while predictions of the Newtonian theory of gravity were not so correct.
 Earlier Einstein considered the gravitational redshift in the approximation of a weak gravitational field and these predictions were subsequently confirmed (see, for example, Ref. \refcite{Pound_60}). Shortly after the creation of GR, one of the first solutions to the theory of gravity
was found by Schwarzschild\cite{Schwarzschild_16} (this solution corresponds to the simplest spherically
symmetric black hole). Under some natural assumptions, a black hole is determined by three parameters, namely, mass $M$, charge $Q$ and angular momentum  $S$ (or spin $a=S/M$), see, for instance book\cite{Chandrasekhar_83} (for astrophysical black holes electric charges must be very small in comparison with their masses since space plasma is quasi-neutral). In 1939, Einstein considered an opportunity for a formation of a Schwarzschild  metric
as a result of the motion along the circles of point masses and in this work
it was concluded that "the Schwarzschild solution cannot arise in physical reality" \cite{Einstein_39}. In 1940s (when there were only a few books on GR)  Einstein's assistant P. Bergman  published his book \cite{Bergman_42} and there he noted with reference
to the Einstein's work that the occurrence of the Schwarzschild solution is very
unlikely. This declaration has been claimed despite the fact that in 1939  the studies by Oppenheimer and Snyder have been published, where numerical modeling shows the possibility of the formation of an astronomical object described by the Schwarzschild solution as a result of dust collapse. Due to the skeptical Einstein's opinion concerning the solution its use for creating models of astronomical
objects was very limited until the appearance of Wheeler's work \cite{Wheeler_68}, where the concept of a "black hole" was used. In turn, the discussion of completely collapsed astronomical objects was motivated by the discovery of pulsars shortly before
\cite{Wheeler_00}.
Here it is important to note that not only successful terminology ({\it black hole} concept is crucial but actually J. A. Wheeler proposed
to use limiting static or stationary in a Kerr case metrics of black holes instead of more complicated dynamical metrics evolving to black hole solutions.
A more detailed discussion of the genesis of the black hole concept is given in papers\cite{Zakharov_02,Zakharov_19}
(where references to works discussing these issues are also given).

Due to significant progress in astronomical facilities and a development of space exploration astronomers began to discuss the possibility of detecting manifestations of GR in the limit of a strong gravitational field. Thus, the work \cite{Fabian_89} shows how the spectral line is distorted, which can be observed in the X-ray range if the emitting region is located near the event horizon of a black hole and subsequently the spectral line of iron $K\alpha$, which has such properties, was detected during the observation of the Seyfert galaxy MCG-6-30-15 by the Japanese ASCA satellite \cite{Tanaka_95} (see also
discussion of these issues\cite{Fabian_00,Jovanovic_12}).  Tanaka et al.\cite{Tanaka_95} showed that the shape of the Fe $K\alpha$ line can be explained
 only in the case emission region for this line must be very close to the event horizon of rapidly rotating Kerr black hole.
 Based on results of our simulations\cite{Zakharov_03} we confirmed these basic statements\cite{Tanaka_95} and
showed that they are rather robust. We also demonstrated an opportunity to evaluate magnetic fields from these observations of Fe  $K\alpha$ lines.\cite{Zakharov_MNRAS03}

Another confirmation of GR predictions in a strong gravitational field approximation has been obtained with observations of gravitational waves from black hole and neutron star mergers.\cite{LIGO_16} Gravitational wave signals were remarkably explained in the framework of GR.\cite{LIGO_16_test} Moreover, constraints on parameters of alternative theories of gravity have been found, for instance, graviton mass has been bounded.\cite{LIGO_21_test}

Shadow observations and remarkable reconstructions of shadows for M87* and Sgr A* by the Event Horizon Telescope (EHT) collaboration are based on three pillars: VLBI facilities, synchrotron radiation and GR analysis of geodesics near black holes.
Really, the VLBI facilities provide an opportunity to observe luminous regions around shadows and these luminous regions arise due to synchrotron radiation while  sizes and shapes of shadows were evaluated from an analysis of photon geodesics near black holes. We should note that
sizes and shapes of shadows do not depend on uncertainties in our knowledge on accretion flows and specific distributions of brightness around shadows.

\section{VLBI, Ground-space VLBI, Radioastron}

Proposals to use very long based interferometer for radio observations were introduced by L. I. Matveenko.\cite{Matveenko_07,Matveenko_07a}
These proposals were met with a great skepticism. Radio astronomers said that such ideas were unrealisable, and astrophysicists said that they did not see astronomical problems for which such a good resolution of radio sources would be necessary.\cite{Matveenko_07}
In the submitted version of the first publication\cite{Matveenko_65} on subject the authors proposed also a ground -- space interferometer for
more precise interferometric observations but this section was removed from the final version of the paper due to demand of editors in this journal.\cite{Matveenko_07} In 1970s Matveenko successfully realized his ideas in joint US -- Soviet experiments.\cite{Matveenko_07}
Using available spacecrafts he proposed a project of a ground -- space interferometer\cite{Matveenko_82} but these ideas were realized
in  HALCA (Highly Advanced Laboratory for Communications and Astronomy)\footnote{Therefore, the first realization of space -- ground interferometer had been done by Japanese astronomers in 1997, see
\url{https://www.isas.jaxa.jp/e/japan_s_history/brief.shtml}.
} also known as VSOP  (VLBI Space Observatory Programme) and Radioastron\footnote{The Radioastron mission was launched in 2011, see
\url{http://www.asc.rssi.ru/radioastron/index.html}.} missions.
A brief description of Matveenko's ideas in a development of VLBI technique was presented.\cite{Zakharov_22}

\section{Synchrotron radiation and its applications in astrophysics}
It was known for more than century that electrons must radiate when they move in magnetic field,\cite{Schott_12} however, in this time there were no experimental facilities to detect these phenomena. In 1939 I. Pomeranchuk studied (the corresponding English translation of his paper was published in 1940\cite{Pomeranchuk_40}) the maximum energy that primary cosmic ray electrons can have on the Earth's surface due to radiation in the Earth's magnetic field.\footnote{The corresponding English translation of his paper was published in 1940\cite{Pomeranchuk_40}. Academician I. Pomeranchuk was an outstanding Soviet theorist and one of the favorite Landau's associates. He was the founder of the theoretical department of the Institute of Theoretical and Experimental Physics founded by A. I. Alikhanov. Brief essays on Pomeranchuk life and his scientific investigations are presented in papers.\cite{Okun_03,Berestetskii_67}}  In 1940s the first betatron accelerators were built\cite{Kerst_42} and Iwanenko and Pomeranchuk applied the Pomeranchuk's expression from paper\cite{Pomeranchuk_40} to estimate the maximal energy attainable in betatron\cite{Ivanenko_44} (a bright representative of the Landau's school B.~Ioffe wrote a book\cite{Ioffe_04} where he presented Pomeranchuk's reminding on the genesis of the paper\cite{Ivanenko_44}). Slightly later, different aspects of magneto-bremsstrahlung  for rapid electrons were analyzed in more extended paper\cite{Artsimovich_45}. Blewett analyzed properties of radiation which could be detected\cite{Blewett_46} and the first detection of X-ray radiation from electrons in 70-MeV synchrotron has been done by Elder at al.\cite{Elder_47} The history of the discovery is presented in papers.\cite{Baldwin_75,Kerst_75,Pollock_82}
Therefore, we state that Pomeranchuk's contribution of  in accelerator physics and its applications is great and as it was noted in paper\cite{Kulipanov_88} he was the  pioneer of synchrotron radiation studies not only in Soviet Union but in the world.

Important contributions in applications of synchrotron radiation to explain emission from  astronomical objects were done by V. L. Ginzburg and I. S. Shklovsky.
Soviet astrophysicist Shklovsky found that synchrotron radiation plays a key role for emission in many astronomical objects, including Sun, Galaxy,
Crab Nebula and many others.\cite{Shklovsky_46} He also noted that an importance of synchrotron emission for many astronomical objects was the most fruitful among all his scientific insights.\cite{Shklovsky_91}
An overview of his astronomical studies with a discussion of  priority issue of astrophysical applications of synchrotron radiation  was presented by V. L. Ginzburg.\cite{Ginzburg_90}

Observations and reconstructions of images around M87* and Sgr A* done by the Event Horizon Telescope Collaboration at 1.3 mm waveband showed that observed emissions in these objects are consistent with a synchrotron radiation model since  strong linear polarizations have been found in there
\cite{Bower_18,EHT_19c,EHT_22a}. In particular, recently there were reported about a discovery of orbital motion of blob moving around the Galactic Center\cite{Wielgus_22}, were signatures if synchrotron radiation were found as well. This discovery was done after an analysis of polarimetric observations with ALMA in April 2017. Therefore, a presence of synchrotron emission sources near these supermassive black holes gives an opportunity to reconstruct darkness (shadows) or in other words we have an opportunity "to see darkness".

\section{Early studies of shadows in  gedanken experiments}

During the creation and development of quantum mechanics and general  relativity, gedanken experiments began to be widely used, the implementation of which can be very difficult or even impossible. Similar thought experiments (and observations) were considered in hypothetical situations during the development of general relativity theory and theory of black holes in particular.
For instance, around 50 years ago C. T. Cunningham and J. M. Bardeen   showed visible shapes of circular orbits of stars moving around extreme Kerr black holes in Fig. 8 in paper,\cite{Cunningham_73}
while  Bardeen showed a shape of dark area\cite{Bardeen_73} (now it is usually called {\it shadow}), in the case if there was a bright screen behind the extreme rotating black hole (a remarkable relativist James Maxwell Bardeen passed away on June 20, 2022 and
in principle he had an opportunity to see reconstructions of his theoretical pictures for Sgr A* and M87*).
         Later, S. Chandrasekhar included a similar picture for shadow in his fundamental book.\cite{Chandra_83}
However, neither Bardeen nor Chandrasekhar discussed an opportunity to detect shadows in observations, perhaps due to three reasons, first, shadow sizes are extremely small for known black hole masses and distances toward them, second, a bright screen behind a black hole looks as an artificial model, third, we have to detect a darkness (shadow) and it is not clear how to distinguish a darkness and a faint area in sky.
In results of his numerical simulations\cite{Luminet_79,Luminet_92} J.-P. Luminet showed  a visible shape of thin accretion disk for a Schwarzschild
black hole and demonstrated that retro-photons form bright rings around shadows, so shadows may be reconstructed from
analysis of rings around them.  D. Holz and J. A. Wheeler considered retro-lensing for photons moving around Schwarzschild black holes and from their analysis we could see for each source even far away from a black hole there is an infinite number of secondary images
just near black hole shadows\cite{Holz_02}. However, these theoretical studies were far beyond the existing observational facilities.

\section{Black holes as dark spots}

 At the edge of the last and current centuries  a creation VLBI network acting at 1.3~mm was under discussion (these ideas were realized by the Event Horison Telescope Collaboration). Angular resolution of this network is around $25~\mu as$\cite{EHT_19a}. In paper\cite{Falcke_00}  Falcke et al. evaluated a discovery potential for such a network in the case of Galactic Center observations and simulated an opportunity to detect shadows with such facilities in the framework of their theoretical model taking into account photon scattering on electrons. The authors concluded
that a shadow could be detected at 1.3~mm while for VLBI network acting at 1.3~cm a shadow detection is hardly ever possible due to photon scattering
at this wavelength. To evaluate the shadow size the authors\cite{Falcke_00} adopted the black hole mass at the Galactic Center as $2.6*10^6 M_\odot$ as it was established in papers\cite{Eckart_96,Ghez_98}.
In papers\cite{Zakharov_05a,Zakharov_05b} we declared that shadows could be detected with mm and sub-mm VLBI ground based networks
and with space-- ground interferometer Millimetron\footnote{https://millimetron.ru/en/general.} which was planning and now it is in a development.
Practically these predictions were realized by the EHT Collaboration for Sgr A*\cite{EHT_22a} and for M87*\cite{EHT_19a} (and these remarkable achievements were results of  advanced observational and computational facilities, new methods of numerical simulations and data analysis and great effort of members of the EHT global collaboration).
Sgr A* shadow diameter should be around $50~\mu as$\cite{Zakharov_05a} if we adopted that the black hole mass is around $4*10^6 M_\odot$ according to papers\cite{Ghez_03,Schodel_03}. For theorists black holes are vacuum solutions of Einstein equations, while for observer black holes are dark spots in the skies. Therefore, simplifying we could say that black holes are dark spots (shadows) and for a general case of accreting matter on astrophysical black hole are formed independently on details of accretion flows. Only for cases of rapidly rotating Kerr black holes  when when emitting matter
is closer to event horizon than photon orbits, the discussed shadows may be not formed. While in cases of rapidly Kerr black holes and truncated disks
shadows are clearly visible as it was shown in paper\cite{James_15} where pictures of thin truncated disks are presented.
Therefore,  we could say that {\it black holes are visible due to dark spots (shadows)} since shadows generally should be formed.
We promoted this approach in a number of publications and talks after publications of papers.\cite{Zakharov_05a,Zakharov_05b}
In addition, we showed that in the case of a Kerr black hole and Bardeen's coordinates $(\alpha, \beta)$ (for vertical and horizontal impact parameters for shadow), for an observer located in the equatorial
plane, the value of the shadow radius in the direction of rotation of the black hole is $\beta(2a)=\sqrt{27}M$ in geometric units (it does not
depend on  the black hole spin $a$), and the shadow
is deformed in a direction parallel to the equatorial
plane; the deformation depends on rotations (these properties of shadows may be easily recognizable in corresponding figures of shadows done by Bardeen\cite{Bardeen_73} and Chandrasekhar\cite{Chandra_83}. Thus, the
shadow diameter for a black hole in the vertical direction  is  $2\sqrt{27}M$ (for an equatorial position of a distant observer).

\section{Shadows for black holes with charge}

The loss cone and shadow evaluation problems are closely connected and sometimes even analytical expressions are the same.
However, since opportunities for experiments are very limited in astronomy, it is very hard to evaluate loss cone in observations while
shadows could be reconstructed.
If we consider Reissner -- Nordstr\"om metric with an electric charge then there are analytical expressions for loss cone of slow moving particles
and photons\cite{Zakharov_91,Zakharov_94}.

Using relations for critical parameters determining a loss cone we could derive relations for the critical impact parameter as a function of electric charge and from obtained relations we have the shadow radius for Schwarzschild black hole is $3\sqrt{3}M$ ($M$ is a black hole mass)
while the shadow radius for the extreme Reissner -- Nordstr\"om is $4M$ and more generally we can see that the shadow radius is a monotonically decreasing function of charge.\cite{Zakharov_05c}
Concerning possible astrophysical applications of Reissner -- Nordstr\"om metric
we did not think that astrophysical black holes have  significant electric charges (in black hole mass units). However, in Randall -- Sundrum theory with extra dimension there is a solution which looks like
a Reissner -- Nordstr\"om where parameter $q=\mathcal{Q}^2$ may be negative ($Q$ is an electric charge). The authors suggested to call this solution
a Reissner -- Nordstr\"om metric with a tidal charge.\cite{Dadhich_00}  Later, it was suggested to apply this solution for the black hole at the Galactic Center to test observational signatures\cite{Bin_Nun_10a,Bin_Nun_10b} but it was shown that in the case of significant negative tidal charge
$q$ the shadow size is so large that it is not consistent with existed constraints on the shadow at Sgr A*\cite{Zakharov_12}.

A derivation of a shadow size as a function of a tidal charge was given in paper.\cite{Zakharov_14}
Here, we recall results given in this paper.
Reissner -- Nordstr\"om  metric can ne written in natural units  ($G=c=1$) as
\begin {equation}
  ds^{2}=-\left(1-\frac{2M}{r}+\frac{\mathcal{Q}^{2}}{r^{2}}\right)dt^{2}+\left(1-\frac{2M}{r}+\frac{\mathcal{Q}^{2}}{r^{2}}\right)^{-1}dr^{2}+
r^{2}(d{\theta}^{2}+{\sin}^{2}\theta d{\phi}^{2}),
\label{RN_0}
\end {equation}
where $M$ is a black hole mass (as we noted earlier), $\mathcal{Q}$ is its charge.
Constants  $E$ and $L$ are connected with photon geodesics, namely $E$ is a photon energy, $L$ is its angular momentum.
If we introduce normalized radial coordinate, impact parameter and charge we have
$\hat {r}=r/M, \xi=L/(ME)$,  $\hat{\mathcal{Q}}=\mathcal{Q}/M.$
Below  we do not write wedge symbol ($\wedge$) under corresponding quantities.
If we introduce new notations $l=\xi^{2}, q=\mathcal{Q}^{2}$, then we have\cite{Zakharov_14}
   \begin {eqnarray}
l_{\rm cr}=\frac{(8q^{2}-36q+27)+\sqrt{D}}{2(1-q)}, \label{RN_D_9}
\end {eqnarray}
where
$D=-512\left(q-\dfrac{9}{8}\right)^3$ and $l_{\rm cr}$ is square of  $\xi_{\rm cr}$ which a critical impact parameter corresponding
to circular photon orbit for a given $q$ parameter.
In the case of a tidal charge\cite{Zakharov_14} (or Horndeski version of scalar-tensor theories\cite{Babichev_17,Zakharov_18}) a parameter $q$ may be negative.


\section{Constraints on black hole charges for M87* and Sgr A*}

Based or results of observations  of M87* in April 2017 and consequent data analysis the EHT collaboration evaluated
the allowed interval for a shadow radius $\theta_{\text{sh~M87*}} \approx 3\sqrt{3}(1
\pm 0.17)\,\theta_{\text{g~M87*}}$ at 68\% confidence level.\cite{Psaltis_20}
Using these constraints  the EHT collaboration found bounds on parameters of
spherically symmetric metrics which could be considered as alternatives to Schwarzschild solution for M87*.\cite{Kocherlakota_21,Wielgus_21}
Among other parameters  the Reissner -- Nordstr\"om charge has been constrained.\cite{Kocherlakota_21}
In our recent paper\cite{Zakharov_22a} we generalized this analysis for a tidal charge case.
Really, similarly to paper\cite{Kocherlakota_21}   it was assumed a presence of the Reissner -- Nordstr\"om with a tidal charge in M87* in paper\cite{Zakharov_22a} and we adopted that
 $\theta_{\text{sh~M87*}} \approx 3\sqrt{3}(1
\pm 0.17)\,\theta_{\text{g~M87*}}$ at  68\% confidence level or $\theta_{\text{sh~M87*}} \in [4.31, 6.08] \theta_{\text{g}~M87*}$, or in other words $\theta_{\text{g}~M87*} \approx 8.1~\mu as$, where $\theta_{\text{g~M87*}}=2M_{M87*}/D_{M87*}$, we have
$q \in [-1.22, 0.814]$ from Eq. (\ref{RN_D_9}). In this case the upper limit of this interval ($q_{upp}=0.814$) corresponds to black hole charge $\mathcal{Q}_{upp}=\sqrt{q_{upp}} \approx  0.902$, shown in Fig.~2 in paper\cite{Kocherlakota_21}.

\begin{figure}[t!]
\begin{center}
\includegraphics[width=0.95\textwidth]{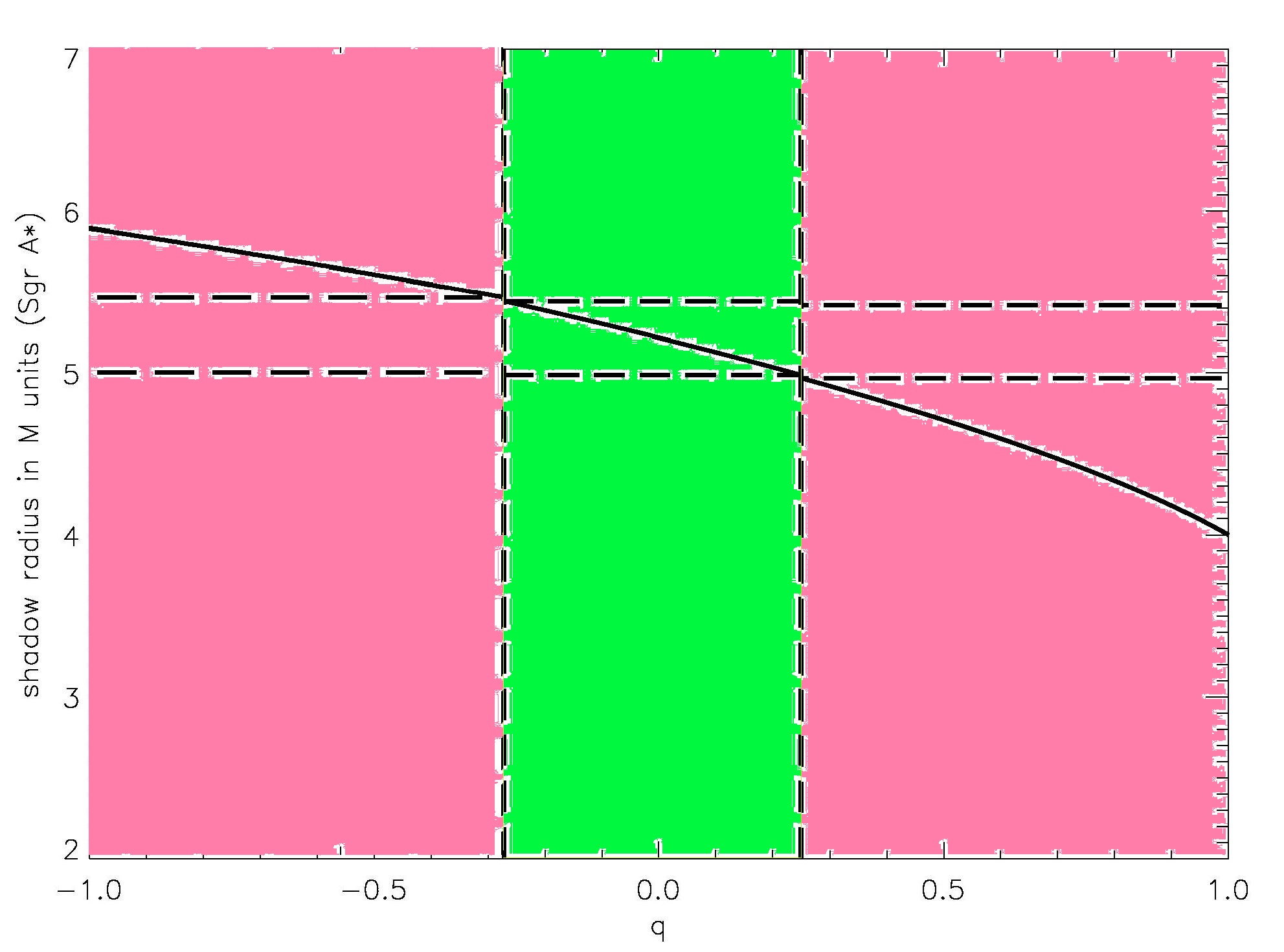}
\end{center}
\caption{Shadow radius $r_{\text{sh~Sgr A*}}$ (solid line)  in $M$ units as a
function of
 $q$. We assume a shadow diameter $\theta_{\text{sh~Sgr A*}} \approx (51.8
\pm 2.3)\mu as $ at 68\% confidence level as it was given by EHT Collaboration\cite{EHT_22a} and in this case, horizontal dashed lines correspond to constraints on shadow radius in $M$ units.
Light green vertical strip corresponds to $q$ parameter, therefore the allowed region for a tidal charge is $-0.27 < q < 0.25$ which are currently consistent with the shadow size estimate done by the EHT collaboration for Sgr A* at 68\% C. L. while rose vertical strips correspond to $q$ parameters which are not consistent with the shadow size estimate.}
 \label{Fig1}
\end{figure}

Similarly, to the procedure to find constraints on a tidal charge for the black hole in M87*, it is possible to constrain a tidal charge
in Sgr A*. We adopt $\theta_{\text{sh~Sgr A*}} \approx (51.8
\pm 2.3)\mu as $ at 68\% confidence level as it was given in paper.\cite{EHT_22a}
Using Eq. (\ref{RN_D_9}) we found constraints on  a tidal charge $-0.27 < q < 0.25$ which is also presented in Fig. \ref{Fig1}.

We should note that Reissner -- Nordstr\"om with a negative tidal charge could be considered as a nice approximation for Kazakov -- Solodukhin black hole\cite{Kazakov_94} since for small "charge" parameter $q_{KS}$ this metric could be approximated by Reissner -- Nordstr\"om with a negative tidal charge. Really, as it was shown\cite{Kazakov_94,Zakharov_23a}
\begin{equation}
  g(r)=-\frac{2M}{r}+\frac{1}{r}\left(r^2-{q_{KS}}^2\right)^{1/2} \approx 1-\frac{2M}{r}-\frac{{q_{KS}}^{2}}{r^{2}},
\label{KS_0}
\end{equation}
where $q_{KS}$ is a Kazakov -- Solodukhin charge. Using this approximation for small $q_{KS}$ we could use Eq. (\ref{RN_D_9}) to find
a shadow radius dependence as a function of charge. In this case we also see that shadow radius dependence is a monotonically increasing function of a Kazakov -- Solodukhin charge.
Constraints on parameters of alternative metrics for the Galactic Center are given in paper\cite{Vagnozzi_22}.

\section{Conclusion}
{\it Shadow} concept passed a long way from theoretical idea introduced by Bardeen and it was transformed in observable (or more precisely
shadow could be reconstructed from observational data) and shadows were reconstructed for M87* and Sgr A* due to a great work done by the EHT Collaboration. Therefore, these remarkable achievements give us an opportunity to check and confirm GR predictions concerning  sizes and shapes of shadows in M87* and Sgr A*. Reconstructions of these shadows could help to constrain parameters of alternative theories of gravity.
Predictions of GR have been checked and confirmed in many experiments and observations\cite{Damour_06, Fabian_15,DePaolis_22} and GR is the best theory of gravity in spite of many proposed alternatives.
Observations of bright stars near the Galactic Center with VLT (GRAVITY) and Keck telescopes
give another opportunity  to test gravitational field there. The leaders of these groups R. Genzel and A. Ghez received Nobel prize in physics in 2020 for a discovery of massive compact object at the Galactic Center (a short review of this studies is presented in paper\cite{Genzel_22}).
Consequences from these observations to create  a suitable model for the Galactic Center
are discussed in papers\cite{Zakharov_22d,Zakharov_22e}.

A few years the GRAVITY collaboration discovered that the Schwarzschild precession of S2 star corresponds to GR estimates \cite{GRAVITY_20}. Recently, constraints on Yukawa gravity parameters have been found from these observational results \cite{Jovanovic_23} (previous estimates of Yukawa gravity paramaters were done in paper\cite{Borka_13}).
Observations of bright stars gave opportunities to evaluate graviton mass as it was done in papers\cite{Zakharov_16,Jovanovic_23b}.

The revision of the paper was prepared in June 2023 when a gravitational community celebrated 135 years since the Alexander Friedmann birth. His cosmological solutions\cite{Friedmann_22,Friedmann_24}
showed us that the Universe expands. However, since a birth of the Universe was not consistent with Soviet philosophy postulates
(which stated that the Universe is infinite in time and space)
Friedmann's solutions were treated
as purely mathematical issues and in Soviet Union in 1930s -- 1950s and it was banned to use these solutions for consideration of Universe evolution.
This ban had a significant negative impact on the development of cosmological and thus on astrophysical research in the USSR\cite{Zakharov_23b}.
In particular, the discovery of cosmic microwave background radiation by T. Shmaonov in 1957 was not interpreted as confirmation of the hot Universe model proposed  G. Gamow and the Shmaonov's  achievement became widely known only after the Nobel Prize was awarded in 1978 to A. Penzias and R. Wilson, who discovered cosmic microwave background radiation in 1965 and in contrast to Shmaonov's results their discovery immediately received the correct cosmological interpretation.
Only in June 1963 when the
Soviet Academy of Sciences celebrated 75 years since the Friedmann birth and the ban was lifted and it was decleared that these results among the most significant and bright achievements of Soviet science\cite{Kapitsa_63,Zeldovich_64}, therefore, we could say that  Soviet physical cosmology was born  around 60 years and active studies in relativistic astrophysics started in Soviet Union. Ya. B. Zeldovich and his school played a decisive role in the rapid and fruitful development of research in this field.

\section*{Acknowledgements}

The author acknowledges the organizers of the ICPPA-2022 for their attention to his contribution for this conference. The author is grateful also to the reviewers for their attentive attitude to the submitted article and valuable remarks.

\end{document}